\documentclass[runningheads]{llncs}
\usepackage{graphicx}
\usepackage{algorithm}
\usepackage{algpseudocode}
\usepackage{amsmath}
\usepackage{xcolor}
\usepackage{amsfonts}
\usepackage{stmaryrd}
\usepackage{amssymb}
\usepackage{pifont}
\MakeRobust{\Call}
\begin{document}
\title{Automating and Mechanizing Cutoff-based Verification of Distributed Protocols}
\titlerunning{Automating Cutoff Proofs}
%
\author{Shreesha G. Bhat \inst{1} \and Kartik Nagar \inst{1}}
\institute{Department of CSE, IIT Madras} 
%
\maketitle              
\begin{abstract}
Distributed protocols are generally parametric and can be executed on a system with any number of nodes, and hence proving their correctness becomes an infinite state verification problem. The most popular approach for verifying distributed protocols is to find an inductive invariant which is strong enough to prove the required safety property. However, finding inductive invariants is known to be notoriously hard, and is especially harder in the context of distributed protocols which are quite complex due to their asynchronous nature. In this work, we investigate an orthogonal cut-off based approach to verifying distributed protocols which sidesteps the problem of finding an inductive invariant, and instead reduces checking correctness to a finite state verification problem. The main idea is to find a finite, fixed protocol instance called the cutoff instance, such that if the cutoff instance is safe, then any protocol instance would also be safe. Previous cutoff based approaches have only been applied to a restricted class of protocols and specifications. We formalize the cutoff approach in the context of a general protocol modeling language (RML), and identify sufficient conditions which can be efficiently encoded in SMT to check whether a given protocol instance is a cutoff instance. Further, we propose a simple static analysis-based algorithm to automatically synthesize a cut-off instance. We have applied our approach successfully on a number of complex distributed protocols, providing the first known cut-off results for many of them.

\end{abstract}
\section{Introduction}

Distributed protocols allow disparate nodes to work together towards completing a task, and form the backbone of today's distributed systems. These protocols are typically specified in a parametric fashion, which means they can be instantiated on a system with any number of nodes. The nodes communicate with each other through message passing, and these messages can be arbitrarily delayed or even lost. However, the distributed protocol is expected to work correctly under all such conditions. Here, correctness is typically defined in terms of a safety property which must be obeyed by every node at every step of the protocol. For example, the safety property of a distributed mutual exclusion protocol would say that two nodes should not be in their critical section at the same time. Since the protocols need to consider every possible network behavior, they are quite complex in nature. Verifying the correctness of distributed protocols then becomes highly important, but this problem is significantly complicated by the parametric nature of the protocol and the asynchronous, non-deterministic nature of the underlying network. Essentially, every possible instantiation of the protocol needs to be proven correct, and each such instantiation itself needs to consider a large number of network behaviors. Further, there could be an infinite number of instantiations of the protocol. 

Recent approaches \cite{FW19,MG19,MP20,PL17,Yao21,Damian19} to verifying distributed protocols typically aim to find an inductive invariant, which is a property of the protocol state satisfied at every step of any protocol instance, is inductive in nature and is stronger than the safety property. However, finding an inductive invariant is very hard, as conceptually, it should encompass all the complex logic that the protocol employs to maintain the safety property \textit{under any abnormal network behavior in any instantiation}. In this work, we consider an alternative cutoff-based approach to protocol verification that cleanly separates the two problems of dealing with \textit{arbitrary instantiations} and \textit{arbitrary network behavior}. We first find a cutoff instance with a fixed, finite number of nodes whose correctness implies the correctness of any arbitrary protocol instance. We then show that the cutoff instance itself is safe in all of its executions. In this way, we only need to consider how the protocol maintains the safety property under arbitrary network behavior in the cutoff instance. Further, since the cutoff instance will have a constant, finite number of nodes, verifying its correctness becomes a finite state verification problem, which can be solved very efficiently.

In this paper, we focus on the problem of finding such a cutoff instance, and automatically showing that it is indeed a cutoff. The definition of a cutoff instance gives us the following characterization: \textit{if there exists a violation of the safety property in any arbitrary protocol instance, then there should also exist a violation in the cutoff instance}. We use this characterization to infer that the cutoff instance should essentially be able to simulate any violating execution in any arbitrary instance. While this still seems like a tall order, we hypothesize that this problem is simpler due to two reasons: (i) a violation of the safety property directly involves only a small number of nodes (for example, a violation of the mutual exclusion property would only require $two$ nodes to be in their critical section together), and further, the participation of other nodes of the system is either not required, or can be simulated by the violating nodes themselves, and (ii) most of the complex logic in the protocol implementation which ensures the absence of a violation can be side-stepped, since we are actually interested in simulating the presence of a violation.

While previous works have also attempted to use cut-off based approaches for verification \cite{RingCutoffs,GSP,cutoffconsensus,bloem_decidability_2015}, they have mostly been limited to either a restricted class of protocols  \cite{GSP} with strong assumptions on the underlying network or a restricted class of specifications \cite{cutoffconsensus}. In this work, we consider a variety of protocols targeting different goals (consensus, mutual exclusion, key-value store, etc.) and do not make any assumptions about the underlying network. Our approach takes as input the protocol description written in the Relational Modeling Language (RML). We first develop a formalization of the cutoff approach which defines sufficient conditions for proving that a given protocol instance is a cutoff instance. This formalization is based on the existence of a simulation relation, which relates the states of any arbitrary protocol instance $L$ and the cutoff instance $C$ at every step of an execution which violates the safety property in $L$. We develop an efficient SMT encoding for checking the correctness of a cutoff instance.

We then use our hypothesis concerning the simplicity of the cutoff instance to develop a static analysis based approach which directly synthesizes the cutoff instance from a violation of the safety property. Beginning from a state which violates the safety property, our analysis moves backwards to identify the necessary protocol actions and state components that could be involved in a violation. We then use the output of the static analysis to create a minimal cutoff instance which faithfully simulates all the protocol actions and state components which could be involved in a violation. Finally, we apply our SMT encoding to check the correctness of the synthesized cutoff instance. We have implemented the proposed approach and applied it on 7 different distributed protocols, providing a cutoff-based proof of correctness for all of them, and successfully synthesizing the cutoff instance automatically for 4 out of the 7 protocols. For the remaining protocols, we only required a small manual change from the output of our synthesis algorithm to obtain the correct cutoff instance and the simulation relation.

To summarize, we make the following contributions:
\begin{enumerate}
	\item We formalize the cutoff approach for distributed protocols written in the RML language, and identify sufficient conditions for proving the correctness of a cutoff instance.
	\item We propose a simple static analysis based approach to automatically synthesize from the protocol description, a cutoff instance and a simulation relation for proving the correctness of the cutoff.  
	\item We have implemented the approach in a prototype tool and have successfully verified on 7 challenging protocols.
\end{enumerate}

The rest of the paper is organized as follows: In \S 2, we illustrate the cut-off based approach to protocol verification and our synthesis algorithm using an example. We formalize the cutoff approach for protocols written in RML in \S 3 and \S 4. Details of our synthesis algorithm are presented in \S 5. Experimental results are given in \S 6, followed by related work and conclusion in \S 7.
\section{Motivating Example: The Sharded Key-Value Store} \label{SectionMotivating}
\subsection{Protocol Description}
As a motivating example to demonstrate our technique, we consider the sharded key-value store protocol described in~\cite{FW19}. The protocol maintains key-value pairs distributed across a set of nodes. A node can \textit{reshard} a key-value pair to any other node by sending a transfer message. The protocol allows messages to be lost or delivered in any order, and implements a re-transmission and acknowledgement logic to ensure that even in the presence of message losses, the key-value pair is reliably resharded to another node. The safety property for this protocol is that no two nodes should ever own a key simultaneously. A detailed pseudocode description of the protocol in the RML language \cite{Ivy} is provided below in Fig. \ref{prot:ShardedKVStore}. 

The protocol is described using a set of sorts, relations and actions. A sort or type is defined for nodes, keys, values and sequence numbers. The relations describe the state of protocol and are defined over these sorts. In a step of the execution, any action can be fired provided that its guard (specified by the \textbf{require} keyword) is satisfied. 

\begin{figure}
\begin{algorithm}[H]
\caption{The Sharded Key Value Store Protocol}
\bgroup
\fontsize{8pt}{10pt}\selectfont
\algrenewcommand\alglinenumber[1]{\fontsize{8pt}{10pt} #1:}
\begin{algorithmic}[1]
\State \textbf{type} $key, value, node, seqnum$
\State 
\State \textbf{relation} $table: node, key, value$
\State \textbf{relation} $transfer\_msg: node, node,
key, value, seqnum$
\State \textbf{relation} $ack\_msg: node, node, seqnum$
\State \textbf{relation} $seqnum\_sent: node, seqnum$
\State \textbf{relation} $unacked: node, node,
key, value, seqnum$
\State \textbf{relation} $seqnum\_recvd: node, node, seqnum$
\State 
\State $owner : value$
\State \textbf{init} $\forall n_1, n_2, k. \: table(n_1, k, owner) \land table(n_2, k, owner) \implies n_1 = n_2$ 
\State 
\State \textbf{action} $reshard (n\_old:node, n\_new:node, k:key, v:value, s:seqnum)$
\State \hspace{\algorithmicindent} \textbf{require} $table(n\_old, k, v) \land \neg seqnum\_sent(s)$
\State \hspace{\algorithmicindent} $seqnum\_sent(s) \gets true$
\State \hspace{\algorithmicindent} $table(n\_old, k, v) \gets false$
\State \hspace{\algorithmicindent} $transfer\_msg(n\_old, n\_new, k, v, s) \gets true$
\State \hspace{\algorithmicindent} $unacked(n\_old, n\_new, k, v, s) \gets true$

\State 
\State \textbf{action} $drop\_transfer\_msg(src:node, dst:node, k:key, v:value, s:seqnum)$
\State \hspace{\algorithmicindent} \textbf{require} $transfer\_msg(src, dst, k, v, s)$
\State \hspace{\algorithmicindent} $transfer\_msg(src, dst, k, v, s) \gets false$
\State 
\State \textbf{action} $retransmit(src:node, dst:node, k:key, v:value, s:seqnum)$
\State \hspace{\algorithmicindent} \textbf{require} $unacked(src, dst, k, v, s)$
\State \hspace{\algorithmicindent} $transfer\_msg(src, dst, k, v, s) \gets true$
\State 
\State \textbf{action} $recv\_transfer\_msg(src:node, dst:node, k:key, v:value, s:seqnum)$
\State \hspace{\algorithmicindent} \textbf{require} $transfer\_msg(src, dst, k, v, s) \land \neg seqnum\_recvd(s)$
\State \hspace{\algorithmicindent} $seqnum\_recvd(s) \gets true$
\State \hspace{\algorithmicindent} $table(dst, k, v) \gets true$
\State 
\State \textbf{action} $send\_ack(src:node, dst:node, k:key, v:value, s:seqnum)$
\State \hspace{\algorithmicindent} \textbf{require} $transfer\_msg(src, dst, k, v, s) \land seqnum\_recvd(s)$
\State \hspace{\algorithmicindent} $ack\_msg(s) \gets true$
\State 
\State \textbf{action} $drop\_ack\_msg(src:node, dst:node, k:key, v:value, s:seqnum)$
\State \hspace{\algorithmicindent} \textbf{require} $ack\_msg(s)$
\State \hspace{\algorithmicindent} $ack\_msg(s) \gets false$
\State 
\State \textbf{action} $recv\_ack\_msg(src:node, dst:node, k:key, v:value, s:seqnum)$
\State \hspace{\algorithmicindent} \textbf{require} $ack\_msg(s)$
\State \hspace{\algorithmicindent} $unacked(src, dst, k, v, s) \gets false$
\State 
\State \textbf{action} $put(n:node, k:key, v:value)$
\State \hspace{\algorithmicindent} \textbf{require} $\exists v'. \: table(n, k, v')$
\State \hspace{\algorithmicindent} $table(n, k, *) \gets false$
\State \hspace{\algorithmicindent} $table(n, k, v) \gets true$
\State 
\State \textbf{safety} $\forall k, n_1, n_2, v_1, v_2, k. \: table(n_1, k, v_1) \land table(n_2, k, v_2) \implies n_1 = n_2 \land v_1 = v_2$
\end{algorithmic}
\egroup
\end{algorithm}
\caption{Sharded Key-value store protocol in RML}
\label{prot:ShardedKVStore}
\end{figure}


The relation $table(N,k,v)$ indicates that the node $N$ holds the key $k$ with the value $v$. A $reshard$ action generates a $transfer\_msg$ from the key's current owner to its new owner. Transfer messages can be arbitrarily dropped (through the $drop\_transfer\_msg$ action), and hence the protocol employs an acknowledgment mechanism, whereby the new owner needs to send an acknowledgment message upon receiving a $transfer\_msg$, and the current owner will keep re-transmitting (through the $retransmit$ action) until it receives an acknowledgment. The acknowledgment message itself can be dropped, requiring re-transmission from the new owner. Finally, messages can be arbitrarily delayed, and it is not possible for a node to know whether its message has been dropped or delayed. Hence, nodes can re-transmit messages an arbitrary number of times until the receipt is acknowledged, resulting in multiple copies of a message floating around. However, since each $transfer\_msg$ message is tagged with a unique sequence number, if a node receives multiple copies of a message, it can ignore repetitions. For example, in the $recv\_transfer\_msg$ action, after receiving an in-flight transfer message and entering the key-value pair in the destination node's table, the sequence number is marked as received in line 30. Now, the guard of the $recv\_transfer\_msg$ action contains a clause which states that the sequence number must not have been received which ensures that older transfer messages do not enter the key value pair into the table of the destination node after the destination node has already received it and potentially re-sharded it to some other node, or changed the associated value with the key through a put action.

\subsection{Cutoff based Verification}

The safety property says that for all runs of the protocol, if there are two table entries for the same key, then these entries must be the same. This property could be violated in two ways: a single node might have two table entries for a key with two different values or there could be two distinct nodes having table entries for the same key. Intuitively, the safety property holds because at all times, either a single node contains the key in its table, or no node contains the key and it is in transit (through a transfer message). Prior approaches to verifying this protocol \cite{FW19} require as input these two distinct phases of the protocol. They then establish inductive invariants for each phase which are strong enough to prove the safety property. These invariants are fairly complex, and ensure for example, uniqueness of transfer messages which have been sent but not received, acknowledgement messages, unacknowledged messages, etc. This highlights the classical problem of crafting inductive invariants: they are significantly more complex than the safety property, and in general encompass almost every detail of the system to block every scenario that can lead to a violation. 

In this work, we consider a radically different approach for verification. Instead of trying to directly prove that a violation of the safety property is not reachable, we instead try to simulate a hypothetical violation occurring in an arbitrary protocol instance through a violation in a fixed, small protocol instance. Then, proving the safety of this fixed, small protocol instance (which is called the cut-off instance) would imply the safety of the protocol. Notice that this completely sidesteps the problem of trying to craft an inductive invariant that ensures unreachability of the violation. Instead, we focus on all possible ways in which a violation can be reached, and then try to simulate such violations in a cutoff instance. 

For the sharded key-value store protocol, we establish that the cutoff in terms of the number of nodes is 2, i.e,  we show that any violation in a system $L$ of any arbitrary size (from hereon, size refers to number of nodes) can be replicated in a cutoff system $C$ with 2 nodes. This cutoff result ensures that model checking the protocol on a two node system is sufficient to establish the correctness of the protocol for a system with an arbitrary number of nodes. This essentially formalizes the `small model' property that has been empirically established by many prior works for bugs in concurrent and distributed systems.

\subsection{Static Analysis}
Our technique employs a static analysis based approach on the protocol description to find out the relevant state components and actions that are necessary for simulating violations of the safety property. Consider a violation in an arbitrary size system $L$ where we have two distinct nodes $a_L$, $b_L$ and key $K$ such that $table(a_L, K, v_1)$ and $table(b_L, K, v_2)$ hold. We are interested in maintaining the state components and simulating the actions that are responsible for this violating state of $L$ in the cutoff system $C$.

At a high level, the static analysis works as follows: We start with the state components that depict the violation. The actions that update these state components are added to the set of relevant actions. However, for these actions to be enabled, their guards will also need to be maintained. So the state components in the guards are also considered. This process of looking at actions that set the newly added state components in the previous iteration and then adding their guard components is repeated until no new entries are added to the set of relevant state components which terminates the process. 

For the sharded key value store protocol, we start with the violating state. Note that, in this state, we are interested in actions and state components responsible for reaching any state that has $table(a_L, K, v_1)$ and $table(b_L, K, v_2)$. We start with this as the initial set of relation entries that we are interested in:
\bgroup 
\fontsize{8pt}{10pt}
$$S = \{ table(a_L, K, v_1), table(b_L, K, v_2) \}$$
\egroup 
Next, we look at actions that set the clauses $table(a_L\langle b_L\rangle, K, v_1\langle v_2 \rangle)$ (we use entries in brackets $\langle \rangle$ to succinctly represent both the clauses). By pattern matching, we can note that any action of the type $put(a_L \langle b_L \rangle ), K, v_1 \langle v_2 \rangle )$  and $recv\_transfer\_msg(*, a_L \langle b_L \rangle, K, v_1 \langle v_2 \rangle), *)$ can set these $table$ entries where $*$ represents that the argument of the action can be any value. These are added to the set of actions being tracked by the algorithm. For these actions to occur, we will also need to add the state components corresponding to the guards of these actions. For the $recv\_transfer\_msg$ actions, the guard contains the clauses $\neg seqnum\_recvd(*)$ and $transfer\_msg(*, a_L \langle b_L \rangle , K, v_1 \langle v_2 \rangle, *)$. For the $put$ actions, we have $\exists v. \: table(a_L \langle b_L \rangle , K, v)$ as the guard clause. For the existential quantifier, we include $table(a_L \langle b_L \rangle , K, *)$ where the value entry is not restricted and therefore all such table entries are tracked as relevant. Therefore, at the end of the first iteration, we have the following set of clauses and action respectively
\bgroup 
\fontsize{8pt}{10pt}
\begin{align*}
S = \{ &table(a_L\langle b_L \rangle, K, v_1\langle v_2\rangle), transfer\_msg(*, a_L \langle b_L \rangle, K, v_1 \langle v_2 \rangle, *), \neg seqnum\_recvd(*)\\
& table(a_L \langle b_L \rangle, K, *)\} \\
A = \{ &put(a_L \langle b_L \rangle ), K, v_1 \langle v_2 \rangle ), recv\_transfer\_msg(*, a_L \langle b_L \rangle, K, v_1 \langle v_2 \rangle), *) \}
\end{align*}
\egroup 
In the second iteration, we look at actions that set the newly added clauses $\neg seqnum\_recvd(*)$, $transfer\_msg(*, a_L \langle b_L \rangle , K, v_1 \langle v_2 \rangle, *)$, $table(a_L \langle b_L \rangle , K, *)$. \\ We then add the clauses corresponding to the guards of these actions.  

In this way, we keep on collecting relevant actions and clauses, terminating in a fixed point after a few iterations. We also simplify the sets by noting that $*$ entries subsume other entries that contain specific values in that field. For example, if the $S$ set contains an entry $table(a_L, K, v_1)$ and also an entry $table(a_L, K, *)$, the latter subsumes the former. On performing such reductions, we get the following sets $S$ and $A$
\bgroup 
\fontsize{8pt}{10pt}
\begin{align*}
S = \{& table(*, K, *), transfer\_msg(*, *, K, *, *), \neg seqnum\_recvd(*), \neg seqnum\_sent(*),  \\ &unacked(*, *, K, *, *)\} \\
A = \{& put(*, K, *), recv\_transfer\_msg(*, *, K, *, *), reshard(*, *, K, *, *) \\& retransmit(*, *, K, *, *)\}
\end{align*}
\egroup

Thus far, we have obtained all possible relevant actions that may lead to the violation and the corresponding clauses required to ensure that these actions fire. There are a number of surprising observations here. The protocol has 8 actions in total, however, the action set we get from the static analysis shows us that only 4 of these actions are actually relevant in a violation. In particular, actions such as $drop\_transfer\_msg$ and $send\_ack$ are not required to simulate a violation. Intuitively, this is because these actions are not necessary to actually transfer a key from one node to another, which is needed for realizing a potential violation. Secondly, although the correctness of the protocol (that is, avoiding a violation) depends on a complex invariant involving uniqueness of a number of state components, we do not require any of that complexity to simulate a violation. The static analysis essentially ignores how exactly a violating state might have been obtained, but instead tries to trace the state components and actions needed that are essential for recreating the violation. For example, it is possible that a transfer message may have been dropped by the network in a violating execution, and hence would need to be re-transmitted. However, the cutoff system need not simulate these steps and may not drop the message in the first place. Intuitively, if a violation occurs in $L$, by maintaining the state components in $S$ and performing only the relevant actions in $A$, we can recreate the violation in the cutoff system $C$. 

\subsection{Simulation Relation \& Lockstep}
While the static analysis gives us the relevant state components and actions that need to be maintained in a cutoff system, we still need to formally prove that any violation in any protocol instance can be simulated by the cutoff instance. To show this, we establish a simulation between any arbitrary instance $L$ and a cutoff instance $C$. The simulation is primarily governed by a \textit{lockstep} which describes the action(s) taken by the cutoff instance $C$ for every action in $L$. An action in $L$ is simulated as zero or more actions in $C$. We also establish a \textit{simulation relation} that holds inductively on the states of both $L$ and $C$ as they progress according to the lockstep. The simulation relation will be strong enough to show that at any step, a violation of the safety property in $L$ will imply a violation in the state of $C$ as well. 

The main ingredients of the simulation relation and lockstep have already been identified via the static analysis, i.e. the relevant state components and corresponding actions required to reach a violating state. What remains is to map the relevant state components and actions of $L$ to corresponding components of $C$. Such a mapping can be obtained by mapping nodes of $L$ to their corresponding simulating node in $C$. Denoting the node mapping as $sim: \mathcal{D}_L \rightarrow \mathcal{D}_C$ (where $\mathcal{D}_x$ represents the set of nodes in the instance $x$), the simulation relation maintains that relevant state components from the set $S$ obtained from static analysis corresponding to any node $n \in \mathcal{D}_L$ in $L$ match the corresponding state component of $sim(n)$ in $C$. The simulation relation does not say anything about the state components which are not relevant for the violation. Similarly, the lockstep ensures that whenever any action from $A$ occurs in $L$, the corresponding action is triggered in $C$. The rest of the actions of $L$ are ignored as they are not relevant to simulate the violation.

Specifically, for the sharded key value store protocol, let us denote the two nodes in the cutoff instance as $a_C$ and $b_C$. Recall that $a_L$ and $b_L$ were nodes of the larger instance $L$ which were involved in the violation. We have $sim(a_L) = a_C$ and $sim(b_L) = b_C$. We map the rest of the nodes to one of $a_C$ or $b_C$, say $b_C$ i.e. $\forall N \in \mathcal{D}_L. \: (N \neq A) \land (N \neq B) \implies sim(N) = b_C$. Intuitively, the node $N_x \in \mathcal{D}_C$ maintains the state and performs the actions $\forall N_L \in \mathcal{D}_L.\: sim(N_L) = N_x$.

Applying the $sim$ mapping on the relevant state components $S$, we get the following simulation relation: 
\bgroup
\fontsize{8pt}{6pt}
\begin{align*}
\forall n, v. \: table_L(n, K, v) &\implies table_C(sim(n), K, v) \\
\forall n_1, n_2, v, s. \: unacked_L(n_1, n_2, K, v, s) &\implies unacked_C(sim(n_1), sim(n_2), K, v, s) \\
\forall  s. \: \neg seqnum\_sent_L(s) &\implies \neg seqnum\_sent_C(s) \\
\forall  s. \: \neg seqnum\_recvd_L(s) &\implies \neg seqnum\_recvd_C(s) \\
\forall n_1, n_2, v, s. \: transfer\_msg_L(n_1, n_2, K, v, s) &\implies transfer\_msg_C(sim(n_1), sim(n_2), K, v,  s)
\end{align*}
\egroup
Here, we use $rel_L$ and $rel_C$ to denote the relation $rel$ of the protocol for the instances $L$ and $C$ respectively. Notice that the simulation relation ensures that any violation of safety property in the protocol state of the larger system (say $table_L(a_L, K, v_1)$ and $table_L(b_L, K, v_2)$) will result in a violation of the cutoff system. The simulation relation is also very simple and straightforward as compared to the inductive invariant required for directly verifying the protocol instance $L$. The lockstep defines the actions fired in the cutoff instance for actions of the larger instance, and ensures that the above simulation relation is maintained for every step of every execution. For actions not in the lockstep, no action is fired in the cutoff instance. Again, the $sim$ mapping and the relevant actions $A$ give the following lockstep:
\bgroup 
\fontsize{8pt}{6pt}
\begin{align*}
&\forall n, v. \: put_L(n, K, c) \textbf{ is simulated as } put_C(sim(N), K, V) \\
&\forall n_1, n_2, v, s. \: reshard_L(n_1, n_2, K, v, s) \textbf{ is simulated as } reshard_C(sim(n_1), sim(n_2), K, v, s) \\
&\forall n_1, n_2, v, s. \: retransmit_L(n_1, n_2, K, v, s) \textbf{ is simulated as } \\& retransmit_C(sim(n_1), sim(n_2), K, v, s) \\
&\forall n_1, n_2, v, s. \: recv\_transfer\_msg_L(n_1, n_2, K, v, s) \textbf{ is simulated as }  \\&recv\_transfer\_msg_C(sim(n_1), sim(n_2), K, v, s) 
\end{align*}
\egroup 

Now, we can show that the simulation relation holds inductively as the two instances $L$ and $C$ execute as-per the lockstep. We reduce this problem to checking satisfiability of a FOL formula where we assume arbitrary protocol states of the two instances $L$ and $C$ such that they obey the simulation relation, but after taking an action according to the lockstep, the protocol states stop obeying the simulation relation. If the resulting formula is unsatisfiable, then the simulation relation is maintained by the lockstep.


\section{Setup}

We consider distributed protocols written in the Relational Modeling Language (RML) \cite{Ivy}. RML is a Turing-complete language, and has been used in many prior works related to distributed protocol verification. RML uses the notions of \textit{relations} and \textit{functions} as used in many-sorted first order logic to describe the state of a distributed protocol. Further, these can be defined over arbitrary domains, as specified by the protocol developer. Constraints on the initial state of the protocol, as well as the safety property can then be directly encoded as FOL formulae over the declared relations and functions. 

The protocol description in RML $\texttt{P} = \langle \texttt{D},  \texttt{R}, \texttt{F}, \Psi, \texttt{A}, \Phi  \rangle$ consists of a set of declarations (\texttt{D},\texttt{R},\texttt{F}), axioms ($\Psi$), actions (\texttt{A}) and a safety property ($\Phi$). The declarations define the vocabulary, i.e. the various domain names (also called $sorts$) \texttt{D} used by the protocol, as well as the relation names \texttt{R} and function names  \texttt{F} along with their signatures over the declared domains. The axioms ($\Psi$) are FOL formulae defined over the vocabulary which encode properties of the domains. $\Phi$ denotes the safety property, which is another FOL formula, while \texttt{A} denotes the actions of the protocol. 

Given the protocol description, we construct a labeled transition system modeling the execution of the protocol. The transition system $\mathcal{A}_{\mathcal{I}}^\texttt{P}=(\Sigma,\Sigma_0,\delta)$ is parameterized by a domain interpretation function $\mathcal{I}$ which associates a finite domain of values with each domain name $\texttt{d} \in \texttt{D}$. For the interpretation function $\mathcal{I}$ to be valid, we require the domains in range of $\mathcal{I}$ to satisfy all the axioms in $\Psi$. Each state $\sigma \in \Sigma$ is an interpretation of function and relation names in $\texttt{F}$ and \texttt{R} to actual functions and relations over the domains defined by the interpretation function $\mathcal{I}$. That is, for a function signature $\texttt{f}:(\texttt{d}_1 \times \ldots \texttt{d}_n) \rightarrow \texttt{d}$ in the protocol description, $\sigma(\texttt{f})$ will be a function of the form $\mathcal{I}(\texttt{d}_1) \times \ldots \mathcal{I}(\texttt{d}_n) \rightarrow \mathcal{I}(\texttt{d})$. Similarly, for a relation $\texttt{r}: (\texttt{d}_1 \times \ldots \texttt{d}_n) \rightarrow \mathbb{B}$, $\sigma(\texttt{r})$ will be an actual relation of the form $\mathcal{I}(\texttt{d}_1) \times \ldots \mathcal{I}(\texttt{d}_n) \rightarrow \mathbb{B}$.

The RML protocol description also consists of a set of axioms $\Psi_0$ constraining the functions and relations in the initial state of the system. We define $\Sigma_0 = \{\sigma \in \Sigma\ |\ \sigma \models \Psi_0\}$ to be the set of states obeying the initialization axioms. Note that the notation $\sigma \models \Psi$ denotes the standard FOL definition of an interpretation ($\sigma$) being the model of an FOL formula ($\Psi$). 

Transitions of $\mathcal{A}_\mathcal{I}^\texttt{P}$ will correspond to actions of the protocol. An action $\texttt{a}(\bar{\texttt{v}}:\bar{\texttt{d}}) = \langle g(\bar{\texttt{v}}), u(\bar{\texttt{v}}) \rangle$ is parameterized over a set of (typed) variable names ($\bar{\texttt{v}}$), and consists of two components: (i) an FOL formula $g$ (also called the guard) defined over the function and relation names of the protocol and which can contain free variables from $\bar{\texttt{v}}$, (ii) an FOL formula $u$ which models the change in the protocol state, defined over unprimed and primed versions of the functions and relations of the protocol. If the current state of the protocol obeys the guard, then the state is updated atomically using the update fomula. The transitions $\mathcal{A}_{\mathcal{I}}^\texttt{P}$ caused by the action \texttt{a} in the protocol are formally defined as follows:
$$
\delta_\texttt{a} = \{(\sigma, \texttt{a}(\bar{x}), \sigma')\ |\ \exists \bar{x} \in \mathcal{I}(\bar{\texttt{d}}).\ \sigma \models g[\bar{x} / \bar{v}] \wedge \sigma,\sigma' \models u[\bar{x} / \bar{v}]   \}\}
$$
That is, for every valuation $\bar{x}$ of the variables $\bar{\texttt{v}}$, there are transitions from states $\sigma$ which obey the guard $g$ to states $\sigma'$ such that $\sigma, \sigma'$ satisfy the update formula. The transition is labeled by the action name along with the actual parameters, i.e. $\texttt{a}(\bar{x})$. The complete set of transitions is obtained by considering the transition set of every action of the protocol: $\delta = \cup_{\texttt{a} \in \texttt{A}} \delta_\texttt{a}$. Let $\delta^*$ denote the reflexive and transitive closure of $\delta$. That is, $\delta^*$ relates states $\sigma$ and $\sigma'$ if there exists a sequence of $n (\geq 0)$ $\delta$ transitions beginning from $\sigma$ and ending in $\sigma'$.

The safety property $\Phi$ is defined as a FOL formulae using the declared domains, functions and relations. In this work, we assume that $\Phi$ only uses universal quantifiers. Hence, $\Phi$ has the form : $\forall (\bar{x}:\bar{\texttt{d}}).\ \phi$. This assumption is consistent with prior works related to distributed protocol verification, and is not restrictive as almost all safety properties can be naturally expressed using just universal quantification.

A trace of $\mathcal{A}_\mathcal{I}^\texttt{P}$ is a sequence of states and transition labels of the form $\sigma_0 a_1 \sigma_1 a_2 \sigma_2 \ldots a_n \sigma_n$ such that $\sigma_0 \in \Sigma_0$ and $(\sigma_{i},a_{i+1}, \sigma_{i+1}) \in \delta$ for all $i, 0 \leq i \leq n-1$. Let $\mathcal{T}(\mathcal{A}_\mathcal{I}^\texttt{P})$ denote the set of traces of $\mathcal{A}_\mathcal{I}^\texttt{P}$. We use $\llbracket \mathcal{A}_\mathcal{I}^\texttt{P} \rrbracket$ to denote the set of reachable states of $\mathcal{A}_\mathcal{I}^\texttt{P}$, i.e. $\llbracket \mathcal{A}_\mathcal{I}^\texttt{P} \rrbracket = \{\sigma'\ |\ \sigma_0 \ldots \sigma' \in \mathcal{T}(\mathcal{A}_\mathcal{L}^\texttt{P}) \}$. A transition system is safe if all of reachable states obey the safety property of the protocol:

\begin{definition}
	Given a distributed protocol $\texttt{P} = \langle \texttt{D},  \texttt{R}, \texttt{F}, \Psi, \Phi, \texttt{A}  \rangle$, a valid interpretation of domains $\mathcal{L}$ obeying $\Psi$, the transition system $\mathcal{A}_\mathcal{L}^\texttt{P}$ is \textbf{safe} if for every reachable state $\sigma \in \llbracket \mathcal{A}_\mathcal{L}^\texttt{P} \rrbracket$, $\sigma \models \Phi$. 	
\end{definition}

While $\mathcal{A}_\mathcal{L}^\texttt{P}$ will be a finite state system (because every domain defined by $\mathcal{L}$ is finite), there can in general be infinite number of domains which satisfy the axioms $\Psi$ of the protocol. For a distributed protocol to be safe, the transition system corresponding to every valid domain interpretation should be safe:

\begin{definition}
	A distributed protocol $\texttt{P} = \langle \texttt{D},  \texttt{R}, \texttt{F}, \Psi, \Phi, \texttt{A}  \rangle$ is safe if for every valid domain interpretation function $\mathcal{L}$ satisfying the axioms $\Psi$, $\mathcal{A}_\mathcal{L}^\texttt{P}$ is safe.
\end{definition} 

\section{Cut-off based Verification}
Each valid interpretation of the domains of a protocol can be seen as a protocol instance. A typical example of a domain with infinite number of valid interpretations is the domain of nodes participating in a protocol. To prove that a protocol is correct, we would need to show its correctness for all possible protocol instances. In cut-off based verification, the idea is to only show correctness for a specific protocol instance called a cut-off instance. In the following, we now formalize cut-off based verification in our framework.

\begin{definition}
	Given a distributed protocol $\texttt{P}$, a \textbf{cut-off instance} $\mathcal{C}$ is a valid interpretation of domains such that if $\mathcal{A}_\mathcal{C}^\texttt{P}$ is safe, then for any valid interpretation $\mathcal{L}$, $\mathcal{A}_\mathcal{L}^\texttt{P}$ is safe.	
\end{definition}

\begin{theorem}
	For a distributed protocol $\texttt{P}$, if $\mathcal{C}$ is a cut-off instance, and $\mathcal{A}_\mathcal{C}^\texttt{P}$ is safe, then the distributed protocol $\texttt{P}$ is safe\footnote{All the proofs can be found in Appendix A}.
\end{theorem}

The above theorem is extremely useful, as it essentially reduces an infinite state verification problem to a finite state verification problem. If we can prove that some protocol instance is a cut-off instance, then we only need to prove the safety of the cutoff instance which is a finite-state system and can be easily verified using model-checking techniques. In this paper, we focus on the harder problems of (i) proving that a protocol instance is a cut-off instance and (ii) automatically synthesizing a cut-off instance. 

Notice that the definition of a cut-off instance implies that if there exists a protocol instance with a violation of the safety property, then the cut-off instance will also have a violation of the safety property. In essence, the cut-off instance can simulate the violation of the safety property in any protocol instance. We use this characterization to propose three conditions which together imply that a protocol instance is a cut-off instance.

These conditions require a simulation relation between states of any arbitrary protocol instance and states of the cut-off instance. Suppose $\mathcal{C}$ is the cut-off instance, resulting in the cut-off transition system $\mathcal{A}_\mathcal{C}^\texttt{P} = (\Sigma^\mathcal{C}, \Sigma_0^\mathcal{C}, \delta_\mathcal{C})$. Let $\mathcal{L}$ be some arbitrary protocol instance, resulting in the system $\mathcal{A}_\mathcal{L}^\texttt{P} = (\Sigma^\mathcal{L}, \Sigma_0^\mathcal{L}, \delta_\mathcal{L})$. To ensure that $\mathcal{C}$ is a cut-off instance, any trace of $\mathcal{A}_\mathcal{L}^\texttt{P}$ which leads to a state violating the safety property should be simulated by a trace of $\mathcal{A}_\mathcal{L}^\texttt{C}$ also leading to a state violating the safety property. Consider a relation $\gamma_\mathcal{L} \subseteq \Sigma^\mathcal{C} \times \Sigma^\mathcal{L}$. We formalize below the conditions which will ensure that $\mathcal{C}$ is a cut-off instance.
\begin{table*}[h]
	\centering
	\begin{tabular}{c | c}
		\hline
		$\varphi_{init}(\gamma_\mathcal{L})$ & $\forall \sigma_\mathcal{L} \in \Sigma_0^\mathcal{L}.\ \exists \sigma_\mathcal{C} \in \Sigma_0^\mathcal{C}.\ (\sigma_\mathcal{L},\sigma_\mathcal{C}) \in \gamma_\mathcal{L}$\\
		\hline
		$\varphi_{step}(\gamma_\mathcal{L})$  & $\forall \sigma_\mathcal{L}, \sigma^{'}_\mathcal{L} \in \Sigma_\mathcal{L}.\ \forall \sigma_\mathcal{C} \in \Sigma_\mathcal{C}.\ \gamma_\mathcal{L}(\sigma_\mathcal{L}, \sigma_\mathcal{C}) \wedge (\sigma_\mathcal{L}, a, \sigma^{'}_\mathcal{L}) \in \delta_\mathcal{L}$ \\ &  $\Rightarrow \exists \sigma_\mathcal{C}^{'} \in \Sigma_\mathcal{C}.\ (\sigma_\mathcal{C}, \sigma_\mathcal{C}^{'}) \in \delta_\mathcal{C}^{*} \wedge \gamma_\mathcal{L}(\sigma_\mathcal{L}^{'}, \sigma_\mathcal{C}^{'}) $\\
		\hline
		$\varphi_{safety}(\gamma_\mathcal{L})$  & $\forall \sigma_\mathcal{L} \in \Sigma_\mathcal{L}.\ \forall \sigma_\mathcal{C} \in \Sigma_\mathcal{C}.\ \gamma_\mathcal{L}(\sigma_\mathcal{L}, \sigma_\mathcal{C}) \wedge \sigma_\mathcal{L} \models \neg \Phi$\\ &  $\Rightarrow \sigma_\mathcal{C} \models \neg \Phi$\\
		\hline
	\end{tabular}
	\caption{Conditions to verify that $\mathcal{C}$ is a cutoff instance.}
	\label{tab:invariants}
\end{table*}

The init condition $\varphi_{init}$ ensures that every initial state of $\mathcal{A}_\mathcal{L}^\texttt{P}$ is related by $\gamma_\mathcal{L}$ to some initial state of $\mathcal{A}_\mathcal{C}^\texttt{P}$. The step condition $\varphi_{step}$ ensures that if states of the protocol instance $\mathcal{L}$ and cut-off instance $\mathcal{C}$ are related by $\gamma_\mathcal{L}$, then after a transition in $\mathcal{A}_\mathcal{L}^\texttt{P}$, the new state of instance $\mathcal{L}$ will continue to be related to a state of $\mathcal{C}$ obtained after 0 or more transitions in $\mathcal{A}_\mathcal{C}^\texttt{P}$. Finally, the safety condition $\varphi_{safety}$ ensures that if a state in $\mathcal{A}_\mathcal{L}^\texttt{P}$ violates the safety property ($\Phi$), then its simulating state in $\mathcal{A}_\mathcal{C}^\texttt{P}$ also violates the safety property. Together, these conditions ensure that any violating trace of any arbitrary protocol instance can be simulated by a violating trace of the cut-off instance.

\begin{theorem}
	Given a distributed protocol \texttt{P} and a valid interpretation $\mathcal{C}$, if for any arbitrary valid interpretation $\mathcal{L}$, there exists a simulation relation $\gamma_\mathcal{L}$ such that $(\varphi_{init} \wedge \varphi_{step} \wedge \varphi_{safety})(\gamma_\mathcal{L})$, then $\mathcal{C}$ is a cut-off instance of \texttt{P}.
\end{theorem}

While the conditions in Table \ref{tab:invariants} ensure that if the cut-off instance is safe, then any arbitrary protocol instance is also safe, we can further refine them based on the following observation: we only need to simulate till the first violation of the safety property, and hence, we can assume that the safety property holds in all states while simulating till the first violation. The refined step condition $\varphi_{step}^{first}$ is defined as follows:

\begin{align*}
   \varphi_{step}^{first}(\gamma_\mathcal{L})  \triangleq &\ \forall \sigma_\mathcal{L}, \sigma^{'}_\mathcal{L} \in \Sigma_\mathcal{L}.\ \forall \sigma_\mathcal{C} \in \Sigma_\mathcal{C}.\ \gamma_\mathcal{L}(\sigma_\mathcal{L}, \sigma_\mathcal{C}) \wedge (\sigma_\mathcal{L}, a, \sigma^{'}_\mathcal{L}) \in \delta_\mathcal{L} \wedge \Phi(\sigma_\mathcal{L})  \\ &  \Rightarrow \exists \sigma_\mathcal{C}^{'} \in \Sigma_\mathcal{C}.\ (\sigma_\mathcal{C}, \sigma_\mathcal{C}^{'}) \in \delta_\mathcal{C}^{*} \wedge \gamma_\mathcal{L}(\sigma_\mathcal{L}^{'}, \sigma_\mathcal{C}^{'})
\end{align*}

\begin{lemma}
	Given a distributed protocol \texttt{P} and a valid interpretation $\mathcal{C}$, if for any arbitrary valid interpretation $\mathcal{L}$, there exists a simulation relation $\gamma_\mathcal{L}$ such that $(\varphi_{init} \wedge \varphi^{first}_{step} \wedge \varphi_{safety})(\gamma_\mathcal{L})$, then $\mathcal{C}$ is a cut-off instance of \texttt{P}.
\end{lemma}

If the protocol is not safe, then we can consider the first violation of the safety property in any arbitrary instance of the protocol. Since the cutoff instance can simulate this first violation, this would imply that the cutoff instance would also not be safe, thus proving the above lemma. We have found in our experiments that the refined conditions are often more effective in proving \textit{cutoff-ness} of a protocol instance. 
\section{Synthesizing the Cutoff Instance}
In this section, we describe our technique to synthesize the cutoff instance and the simulation relation from the protocol description. 

\subsection{Pre-processing \& Notation}
Given the protocol description, we perform a pre-processing step to aid with the static analysis and synthesis procedure. Formally, the protocol description in RML is statically pre-processed to obtain a metadata structure $P$ which has actions, relations, sorts and functions denoted by $P.actions$, $P.sorts$, $P.relations$, $P.functions$. 
    
A relation $r \in P.relations$ is associated with the metadata $r.args$ which is an list of sorts from $P.sorts$ representing the types of its arguments. We use $r.out$ to denote the sort of the output which will be boolean ($\mathbb{B}$) for relations. A function $f \in P.functions$ is associated with $f.args$ which is list of arguments from $P.sorts$, and $f.out \in P.sorts$ which represents the type of the output of the function.

An action $a \in P.actions$ has the following metadata: $a.named\_arguments$ is a list of all the argument names of the action. For each relation or function present in the guard, we create a tuple $(x, l, o) \in a.guard\_atoms$ where, 
\begin{itemize}
    \item $x$ is either a relation $r \in P.relations$ or a function $f \in P.functions$.
    \item $l$ is the list of argument names to $x$ in the guard. Each argument name must either a member of $a.named\_arguments$ or $*$ which represents that the argument was another function or relation.
    \item $o$ would be either a value of type boolean or $f.out$, or $*$. If $r(l)$ (or $f(l)$) is statically required by the guard to be a constant of type $x.out$, then $o$ would be that constant, else, $o$ is set to $*$.
\end{itemize}
Further, action $a$ is also associated with $a.body$: for each relation or function that is updated in the action, we create a tuple $(x, l, o)$ where $x, l, o$ are the same as defined earlier for the guard, except that $o$ now represents what state $x(l)$ will be in after executing the action. Each member of $l$ is again either a member of $a.named\_arguments$ or $*$. The field $o$ can similarly contain $*$, or a constant. In all of the above cases, the $*$ value indicates that the actual value cannot be statically determined. 

As an example, we refer back to an action from the Sharded Key Value store example from \S2. Consider the action $reshard$ action. The above pre-processing steps would yield the following attributes for the $reshard$ action
\begin{align*}
    reshard.named\_arguments = [&n\_old, n\_new, k, v, s] \\
    reshard.guard\_atoms = [& (table, [n\_old, k, v], true) \\
                            & (seqnum\_sent, [s], false) ] \\
    reshard.body = [& (seqnum\_sent, [s], true) \\
                    & (table, [n\_old, k, v], false) \\
                    & (transfer\_msg, [n\_old, n\_new, k, v, s], true) \\
                    & (unacked, [n\_old, n\_new, k, v, s], true) ] 
\end{align*}

We define a few more terms that are used in the algorithm description:

\begin{itemize}
    \item An \textit{instantiation} of $a.named\_arguments$ for an action $a$ is defined to be a concrete assignment of the named arguments to a value of their appropriate sort. Formally, an instantiation is a map from keys being the named arguments of the action and the values being their assignments. A value of $*$ represents that the corresponding named argument can take any value.
    \item An \textit{action invocation} is defined as a tuple $(a, I)$ where $a \in P.actions$ and $I$ is an instantiation of $a.named\_arguments$. 
    \item We define a \textit{clause} as a triple $(x, l, o)$ where $x \in P.relations \cup P.functions$, $L$ is a list of values (some of which can be $*$) conforming to the types in $x.args$ and $o$ is either a constant of type $x.out$ or $*$. 
\end{itemize}

Referring back to our motivating example, an instantiation of the named arguments of the $reshard$ action would be $$I = [n\_old: *, n\_new: a_L, k: K, v: *, s: *]$$ and correspondingly, an action invocation would be the tuple $(reshard, I)$. Similarly, a clause on the $table$ relation would be $(table, [a_L, K, *], true)$. 
\subsection{Static Analysis}
The static analysis algorithm takes as input the protocol metadata structure $P$ and an initial set of clauses $S_{init}$. $S_{init}$ will be derived from the safety property of the protocol; more details are provided in \S 5.3. We maintain two sets $S$ and $A$ where $S$ contains a set of clauses and $A$ a set of action invocations. In each iteration, we consider all the new clauses added to the set $S$ in the previous iteration (line~\ref{SA:eachclause}). For each clause $c$, in line~\ref{SA:actionscall}, we invoke $\Call{ActionsThatSet}{P, c}$ to obtain all the action invocations that potentially set the clause $c$. We then add the guards for all these action invocations to the set $S$ in line~\ref{SA:guardscall}. The while loop at line~\ref{SA:loop} terminates when no new clauses have been added in the previous iteration, thus indicating that we have reached a fixed point. 
\begin{algorithm}[H]
\textbf{Arguments}: $P$ the program, $S_{init}$ a set of clauses \\
\textbf{Returns}: $S$ a set of clauses, $A$ a set of action invocations
\caption{\Call{StaticAnalysis}{}} \label{alg:StaticAnalysis}
\begin{algorithmic}[1]

\Procedure{StaticAnalysis}{$P, S_{init}$} 
\State $S \gets S_{init}$
\State $S\_{prev} \gets \emptyset$
\State $A \gets \emptyset$
\While{$S \neq S_{prev}$} \Comment{Loop till a fixed point is obtained} \label{SA:loop}
    \State $S_{prev} \gets S$
    \For{each clause $c$ in $S \setminus S_{prev}$} \Comment{For each new clause} \label{SA:eachclause}
        \State $A_{t} \gets \Call{ActionsThatSet}{P, c}$ \label{SA:actionscall}
        \For{each action invocation $act$ in $A_t$} 
            \State $S \gets S \cup \Call{GuardsFor}{P, act}$ \label{SA:guardscall}
        \EndFor        
        \State $A \gets A \cup A_t$
    \EndFor
\EndWhile
\State \Return{$S, A$}
\EndProcedure
\end{algorithmic}
\end{algorithm}

\begin{algorithm}[H]
\textbf{Arguments}: $P$ the program, and a clause $c$ \\
\textbf{Returns}: $A$ a set of action invocations
\caption{\Call{ActionsThatSet}{}}
\begin{algorithmic}[1]
\Procedure{ActionsThatSet}{$P, c$}
\State $A = \emptyset$
\For{$a \in P.actions$}
    \For{$at\_update = (x, l, o)$ in $a.body$}
        \If{$at\_update.x == c.x$} \label{ATS:checkfr}
            \If{$\neg$ ($c.o \neq *$ and $at\_update.o \neq *$ and $c.o \neq at\_update.o$)} \label{ATS:compatibility}
                \State Create an instantiation $I$ of $a.named\_arguments$, initialized to $*$\label{ATS:inst}
                \If{$\Call{PatternMatch}{at\_update.l, c.L}$} 
                    \For{$i \in 1, len(at\_update.l)$ if $at\_update.l[i] \neq *$} \label{ATS:for1}
                        \State $I[at\_update.l[i]] \gets c.L[i]$ \label{ATS:for2}
                    \EndFor
                    \State $r \gets (a, I)$ 
                    \State $A \gets A \cup \{r\}$
                \EndIf
            \EndIf
        \EndIf
    \EndFor
\EndFor
\State \Return{$A$}
\EndProcedure
\end{algorithmic}
\end{algorithm}
The function $\Call{ActionsThatSet}{P, c}$ takes as input the program $P$ and a clause $c$ to return a set of action invocations $A$ which potentially set the clause $c$. The algorithm works by pattern matching. We iterate over actions and for each atomic update in the body of the action, we check if the atomic update tuple matches the tuple in the clause with respect to the function/relation it updates in line~\ref{ATS:checkfr}. The if condition in line~\ref{ATS:compatibility} fails only if both the atomic update output and the clause output can be determined statically and they do not match each other. 

As an example, assume that the if condition in line~\ref{ATS:checkfr} passes i.e. both the atomic update and the clause refer to the same function/relation $x$ i.e. $c.x = at\_update.x = x$. If the clause output $c.o = *$ and $at\_update.o = true$ then this means that we are interested in actions that potentially affect $x(c.L)$ in any way, and this atomic update therefore satisfies that requirement. Similarly, if $c.o = true$ and $at\_update = *$, this means that we are interested in actions that set $x(c.L) = true$, but the value that the atomic update alters $x(at\_update.l)$ cannot be determined statically. Therefore, conservatively, we assume that the atomic update could potentially alter it as required. But, if $c.o = true$ and $at\_update = false$, then the if condition fails as the outputs can be determined statically but do not match.

In line~\ref{ATS:inst} we create an instantiation of $a.named\_arguments$ initialized to *. The $\Call{PatternMatch}{}$ function considers the arguments of the update atom and the clause atom $at\_update.l, c.L$ and checks for inconsistencies. For example, $at\_update.l=(a, b, a)$ and $c.L=(1, 2, *)$ would pass the check whereas $at\_update.l=(a, b, a)$ and $c.L=(1, 2, 3)$ would fail the check. If the pattern match succeeds, the for loop instantiaties the named arguments in $at\_update.l$ based on $c.L$. The tuple $(a, I)$ now forms the action invocation which is added to the set of action invocations returned by the algorithm.
\begin{algorithm}[H]

\textbf{Arguments}: $P$ the program, an action invocation $act$ \\
\textbf{Returns}: $G$ a set of clauses
\caption{\Call{GuardsFor}{}}
\begin{algorithmic}[1]
\Procedure{GuardsFor}{$P, act$}
\State $G = \emptyset$
\For{$g = (x, l, o) \in a.guards$} \label{GF:guards}
    \State Create a list $L$ of length $g.l$, initialized to $*$
    \For{$i \in 1, len(g.l)$ if $g.l[i] \neq *$} \label{GF:for1}
        \State $L[i] \gets act.I[g.l[i]]$ \label{GF:for2}
    \EndFor
    \State $G \gets G \cup \{(g.x, L, g.o)\}$ \label{GF:clausetuple}
\EndFor
\State \Return{$G$}
\EndProcedure
\end{algorithmic}
\end{algorithm}
The $\Call{GuardsFor}{}$ actions returns the set of clauses involved in the guard for an action invocation. We iterate through all the guard atoms of the action in line~\ref{GF:guards}. The for loop in lines~\ref{GF:for1}-\ref{GF:for2} assigns concrete values to the named arguments in $g.l$ using the instantiation $I$ provided in the action invocation. Then a clause tuple is created in line~\ref{GF:clausetuple} and added to the list of clauses returned by the algorithm. 

As an example to demonstrate how $\Call{GuardsFor}{}$ and $\Call{ActionsThatSet}{}$ work, we refer back to Sharded Key Value store example considered in \S2. 

If $\Call{ActionsThatSet}{P, (unacked, [*, a_L, K, *, *], true)}$ is invoked, when iterating over the $reshard$ action, the action invocation $(reshard, [n\_old: *, n\_new: a_L, k: K, v: *, s: *])$ would be added to the set $A$ by the algorithm. Similarly, if $\Call{GuardsFor}{reshard, [n\_old: *, n\_new: a_L, k: K, v: *, s: *]}$ is invoked, the set 
$$G = \{ (seqnum\_sent, [*], false), (table, [*, K, *], true) \}$$
would be returned.

\subsection{Synthesizing the Cutoff Instance, Simulation Relation \& Lockstep}

\textbf{Cutoff Instance.} We start with the safety property $\Phi$ in the RML description. As described in \S3, the safety property only contains universal quantifiers and hence is as a formula of the form 
$$ \forall (\bar{x}:\bar{\texttt{d}}).\ \phi$$
The size of the cutoff system is defined to be the number of universally quantified nodes in the safety property.

\noindent\textbf{Obtaining $S_{init}$.} Consider any arbitrary size instance $L$ with $\mathcal{D}_L$ denoting the set of nodes. To begin with the static analysis, we need to provide an initial set of clauses $S_{init}$ as input along with the pre-processed protocol metadata structure $P$. To obtain $S_{init}$, we first negate the safety property and instantiate all the existentially quantified variables. We define $\mathcal{D}_L^{v} \subseteq \mathcal{D}_L$ the set of instantiated nodes or \textit{violating nodes}. We then process the resulting FOL formula $\neg \phi$ to obtain the set of clauses involved in the formula. This processing is similar to the pre-processing performed on the FOL formula representing the guards and body of the actions to obtain guard atoms and atomic updates.

As an example, consider the safety property for the Sharded Key Value store protocol from \S2. We have $$ \forall N_1, N_2, K, V_1, V_2. \: table(N_1, K, V_1) \land table(N_2, K, V_2) \implies N_1 = N_2 \land V_1 = V_2 $$
As there are 2 quantifiers on nodes, the cutoff for the protocol is 2. Negating and instantiating $N_1 = a_L, N_2 = b_L, K = k, V_1 = v_1$ and $V_2 = v_2$, we get
$$table(a_L, k, v_1) \land table(b_L, k, v_2) \land (n_1 \neq n_2 \lor v_1 \neq v_2) $$
giving us the following set of clauses after processing
$$\{ (table, [a_L, k, v_1], true), (table, [b_L, k, v_2], true)) \}$$

\noindent\textbf{Synthesizing the Simulation Relation and Lockstep.} Having obtained $S_{init}$, we can now invoke $\Call{StaticAnalysis}{P, S_{init}}$ to get the set of clauses $S$ and set of action invocations $A$. We also have the cutoff instance $C$ with $\mathcal{D}_C$ set of nodes. To define the lockstep and simulation relation, we map the nodes of the violating instance to nodes of the cutoff system. Such a mapping $sim: \mathcal{D}_L \rightarrow \mathcal{D}_C$ is defined as follows. Firstly, by construction, $|\mathcal{D}_L^v| = |\mathcal{D}_C|$ i.e., the number of nodes involved in the violation is the same as the number of nodes in the cutoff system. Consequently, we perform a one-to-one mapping of nodes from $\mathcal{D}_L^v$ to $\mathcal{D}_C$. For the rest of the nodes $\mathcal{D}_L \setminus \mathcal{D}_L^v$ in the system $L$, we make the following observations
\begin{itemize}
    \item If $S$ and $A$ obtained from the static analysis do not have any components containing $*$ in any field of the node type, this implies that only actions and state components of the violating nodes are sufficient to simulate the violation. In such a case, there is no need to map nodes from  $\mathcal{D}_L \setminus \mathcal{D}_L^v$ as they will never appear in the simulation relation or lockstep. 
    \item If $S$ or $A$ obtained from the static analysis has components containing $*$ in any field of the node type, we map all the nodes from $\mathcal{D}_L \setminus \mathcal{D}_L^v$ to one of the nodes in $\mathcal{D}_C$. 
\end{itemize}
Intuitively, the simulation relation states that for all the clauses that are relevant to the violation (as obtained by the static analysis procedure) in the larger system $L$, the same state components are maintained in the cutoff system but in the relation tables of the simulating nodes (as per the $sim$ mapping). Similarly, the lockstep states that the relevant actions are performed in the cutoff system, but by the simulating nodes. 

Given $S, A$ and $sim$, we obtain the simulation relation and lockstep using the procedure $\Call{SimAndLockstep}{S, A, sim}$ in Algorithm \ref{alg:sim}. The procedure returns the simulation relation $\gamma$ as a FOL fomula and the lockstep $\tau$ as an abstract map from action invocations of the larger system to action invocations of the cutoff system.  
\begin{algorithm}[H] 
\textbf{Arguments}: Set of clauses $S$, action invocations $A$ and mapping $sim: \mathcal{D}_L \rightarrow \mathcal{D}_C$ \\
\textbf{Returns}: FOL formula $\gamma$ representing the simulation relation and lockstep $\tau$ as a map from actions of the larger system to actions of the cutoff system
\caption{Function to obtain simulation relation and lockstep}
\begin{algorithmic}[1]
\Procedure{SimAndLockstep}{$S, A, sim$}
\State $\gamma \gets true$
\For{each clause $c \in S$}
    \State Assign unique variable names $\bar{v}$ for all entries containing $*$ in $c.L$
    \State Replace the $*$'s in $c.L$ with assigned variable names to get $\mathcal{L}_{args}$
    \State Replace each node variable $n$ in $\mathcal{L}_{args}$ with $sim(n)$ to get $\mathcal{C}_{args}$
    \If{$c.o == *$} 
        \State \Comment{In this case, we assert that the function/relation entries are equal in the larger system and cutoff system}
        \State $\gamma \gets \gamma \bigwedge  (\forall \bar{v}. \: c.x(\mathcal{L}_{args})  = c.x(\mathcal{C}_{args}) $)
    \Else 
        \State \Comment{In this case, we assert that if the relation/function entry takes the value $c.o$ in $L$, it also does so in $C$}
        \If{$x.out$ is of node type}
            \State $\gamma \gets  \gamma \bigwedge (\forall \bar{v}. \: (c.x(\mathcal{L}_{args}) = c.o ) \implies  (c.x(\mathcal{C}_{args}) = sim(c.o))$)
        \Else
            \State $\gamma \gets  \gamma \bigwedge (\forall \bar{v}. \: (c.x(\mathcal{L}_{args}) = c.o ) \implies  (c.x(\mathcal{C}_{args}) = c.o)$)
        \EndIf 
    \EndIf 
\EndFor 
\State Initialize an empty map $\tau$
\For{each action invocation $act \in A$}
    \State Assign unique variable names $\bar{v}$ for all keys containing $*$ in $act.I$ 
    \State Replace values containing $*$'s with the corresponding variable from $\bar{v}$ in $act.I$ to get $act_L.I$
    \State Replace each node value $n$ in $act_L.I$ with $sim(n)$ to get $act_C.I$ 
    \State Define $act_L = (act.a, act_L.I)$ and $act_C = (act.a, act_C.I)$
    \State $\forall \bar{v}. \: \tau(act_L) \gets act_I$
\EndFor 
\State \Return{$\gamma, \tau$}
\EndProcedure

\end{algorithmic}
\label{alg:sim}
\end{algorithm}
\noindent\textbf{Cutoff Verification}: To prove that the synthesized cutoff instance is actually a cutoff for the protocol, we need to generate first order formulas for each of the 3 properties $\varphi_{init}(\gamma_{\mathcal{L}}), \varphi_{step}^{first}(\gamma_{\mathcal{L}})$ and $\varphi_{safety}(\gamma_{\mathcal{L}})$ mentioned in \S3. With the simulation relation $\gamma$ synthesized earlier, we can directly use the same to generate the encodings. Furthermore, for $\varphi_{step}^{first}(\gamma_{\mathcal{L}})$, we remove the existential quantifier over the state $\sigma_{\mathcal{C}}^{'}$ after the transition by providing a candidate transition in the system $\mathcal{C}$ as per the lockstep. The new formula $\varphi_{step}^{first}(\gamma_{\mathcal{L}}, \tau_\mathcal{L})$ becomes, 

\begin{align*}
   \varphi_{step}^{first}(\gamma_\mathcal{L}, \tau_\mathcal{L})  \triangleq &\ \forall \sigma_\mathcal{L}, \sigma^{'}_\mathcal{L} \in \Sigma_\mathcal{L}.\ \forall \sigma_\mathcal{C} \in \Sigma_\mathcal{C}.\ \gamma_\mathcal{L}(\sigma_\mathcal{L}, \sigma_\mathcal{C}) \wedge (\sigma_\mathcal{L}, a, \sigma^{'}_\mathcal{L}) \in \delta_\mathcal{L} \wedge \Phi(\sigma_\mathcal{L}) \\ & \wedge (\sigma_\mathcal{C}, \tau_\mathcal{L}(a), \sigma^{'}_\mathcal{C}) \in \delta_\mathcal{C}    \Rightarrow \gamma_\mathcal{L}(\sigma_\mathcal{L}^{'}, \sigma_\mathcal{C}^{'})
\end{align*}


\section{Experimental Results}
We have applied the proposed strategy on a variety of different distributed protocols, given in Table \ref{tab:Results}. Our technique works in two parts, where we first attempt to automatically synthesize the cut-off instance, and then attempt to prove its correctness. For proving correctness of a cutoff instance, we generate a FOL encoding of the 3 conditions $\varphi_{init}(\gamma_{\mathcal{L}}), \varphi_{step}^{first}(\gamma_{\mathcal{L}}, \tau_{\mathcal{L}})$ and $\varphi_{safety}(\gamma_{\mathcal{L}})$. We reduce the problem of checking correctness to satisfiability of the generated FOL formulae. For example, for checking the $\varphi_{step}^{first}(\gamma_{\mathcal{L}}, \tau_{\mathcal{L}})$ condition which is a condition of the type $p \implies q$ to be correct, we check whether $p \land \neg q$ is unsatisfiable. We use Z3~\cite{Z3Solver} as our backend SMT solver. Table \ref{tab:Results} summarizes our experimental results. Notice that the time taken for each protocol is in the order of few milliseconds.

\begin{table}[]
    \centering
    \begin{tabular}{|c|c|c|c|c|c|}
        \hline  
        Protocol & Cutoff Size & Time Taken(ms) & $|\gamma|$ & $|\tau|$/total actions &  Automated Synthesis \\
        \hline 
        Sharded Key Value Store\cite{Ironfleet} & 2& 20& 5& 4/8& \ding{51} \\
        \hline
        Leader Election in a Ring\cite{LERing} & 2& 30 & 2& 2/2& \ding{51} \\
        \hline
        Centralized Lock Server\cite{Wilcox2015VerdiAF} & 2& 40 & 5& 5/5& \ding{51} \\
        \hline
        Ricart Agrawala\cite{RicartAgrawala} & 2& 30& 6& 4/4$^{*}$ &  \ding{51} \\
        \hline
        Basic Key Value Store\cite{FW19} & 2& 20& 2& 2/2& \ding{55} \\
        \hline
        Two Phase Commit\cite{Gray1978} & 3& 40& 21& 7/7& \ding{55} \\
        \hline
        Distributed Lock Server\cite{Ivy} & 2& 30& 5& 2/2& \ding{55} \\
        \hline 
    \end{tabular}
    \caption{$\gamma$ is a FOL formula of the type $\bigwedge_{i = 1}^{| \gamma|} (p \implies q)$ therefore $|\gamma|$ represents the number of clauses of the type $p \implies q$ in the simulation relation. $|\tau|$ refers to the number of actions that are simulated in the cutoff system. Time taken refers to the total time taken by our synthesis+verification procedure. \\
    \small{*only actions involving the violating nodes are simulated}}
    \label{tab:Results}
\end{table}

\subsection{Protocols}
We now present a detailed description of each protocol and its cutoff instance

\noindent\textbf{Leader Election in a Ring}: We considered the Leader Election in a ring protocol as presented in \cite{LERing}. The protocol deals with electing a unique leader in a ring setting. Each node has a unique ID and there exists a total order on the IDs of the nodes. The protocol defines static relations $between$ (to define a notion of a node being between two nodes) and $next$ (to define the clockwise neighbouring node for every node) which are used to describe the ring topology. The protocol works as follows, each node sends a message containing its ID to its clockwise neighbour. When any node receives an ID, the ID is forwarded to its neighbour only if the ID in the message is greater than its own. When a node receives its own ID, it elects itself as a $leader$. The safety property describes that no two nodes are ever elected as leaders i.e. $leader(N_1) \land leader(N_2) \implies false$. Intuitively, the protocol works correctly because we have a total order on the IDs and therefore only the node with the highest ID is elected as the leader. 

To synthesize a cutoff instance, we instantiate a violation in an arbitrary instance containing two nodes $a_L$ and $b_L$ such that $leader(a_L)$ and $leader(b_L)$ both hold. We then synthesize a cutoff instance with 2 nodes $a_C$ and $b_C$ such that $ID(a_L) = ID(a_C)$ and $ID(b_L) = ID(b_C)$. With the $sim$ relation mapping $a_L \rightarrow a_C$ and every other node to $b_C$ along with the $S$ and $A$ sets obtained from the lockstep, we are able to generate a correct cutoff instance.

\noindent\textbf{Ricart-Agrawala}: We consider the Ricart Agrawala protocol for distributed mutual exclusion as presented in \cite{RicartAgrawala}. A node sends request messages to all other nodes to enter the critical section. A responder node replies to a requester only if the responder is not holding the lock and hasn't received nor sent a reply to/from the requester earlier. To enter the critical section, a node must have received a reply from every other node. To leave the critical section, a node simply discards previous replies and relinquishes control of the lock. The safety property is mutual exclusion i.e. no two nodes will ever enter the critical section/hold the lock simultaneously. 

In synthesizing the cutoff instance, we start with a violation in an arbitrary instance where two nodes $a_L$ and $b_L$ held the lock simultaneously. The resulting static analysis yielded sets $A$ and $S$ where only the actions of the two violating nodes and their state components were relevant to simulate the violation. We thus simulate a cutoff instance with two nodes $a_C$ and $b_C$ with $sim(a_L) = a_C$ and $sim(b_L) = b_C$. Note that there is no need to simulate actions of other nodes as they do not appear in the backward analysis. This is noteworthy, as for other protocols, the backward analysis eventually results in inclusion of actions and state components of non-violating nodes as well. 

\noindent\textbf{Centralized Lock Server}: The Lock Server protocol~\cite{Wilcox2015VerdiAF}, implements a centralized lock service. Clients request for the lock by sending a $lock$ message. Similarly, a client can relinquish the lock (if it owns it) by sending an $unlock$ message. The protocol allows for the client to send arbitrary number of duplicate $lock$ messages. The network picks a message at random to be handled by the server. The server maintains a queue of client nodes that have requested for the lock while granting the lock the node at the head of the queue. To handle an $unlock$ message, the server removes the node from the head of the queue and sends a $grant$ message to the new head (if any). On receiving a $grant$ message, the receiving client node holds the lock. The safety property is again mutual exclusion i.e. no two nodes hold the lock simultaneously. 

We implemented the protocol in RML by emulating the network semantics. We implement a unique sequence number for every $lock$, $unlock$ and $grant$ message. The functionality of picking a random network message is then equivalent to firing a $handle$ action for one of the previously un-handled sequence numbers at random. We also implement the queue as a triple $(head, tail, q)$ where $head$ and $tail$ are integers and $q: \mathbb{Z} \rightarrow node$ is a function that outputs the queue elements for valid indices between $head$ and $tail$ inclusive. 

To synthesize the cutoff instance, we instantiate a violation with two nodes $a_L$ and $b_L$ that both hold the lock. In the cutoff system, we have two nodes $a_C$ and $b_C$. The static analysis requires us to simulate the actions of nodes apart from the violating nodes, hence we map $sim(a_L) = a_C$ and the rest of the nodes to $b_C$. At a high level, the $lock, unlock$ and $grant$ messages of all nodes that map to $b_C$ are sent and handled by the node $b_C$ in the cutoff system. The simulation relation also states that the queue in the cutoff system is identical to the queue of the larger system with the $sim$ mapping applied to the contents of the queue. \\

\noindent For the following protocols, our synthesis strategy was not able to directly generate the correct cutoff instance. We used the output of our synthesis algorithm, and performed a few manual tweaks to obtain the correct cut-off instance.

\noindent\textbf{Two Phase Commit}: We consider the Two-Phase Commit protocol as provided in \cite{Gray1978,Yao21} for performing atomic commits in a distributed setting. Nodes can decide to $vote\_yes$, $vote\_no$ locally. Nodes can also fail arbitrarily as emulated by the $fail$ action (which marks the node as $\neg alive$). The global actions $go\_commit$ is triggered if all the nodes have voted yes, whereas the $go\_abort$ action is triggered if there exists a node that has failed or voted no. After the global actions $go\_commit$ or $go\_abort$ are triggered, the nodes can locally $decide\_commit$ or $decide\_abort$ respectively. We consider the key safety property that states that $decide\_commit(N_1) \land decide\_abort(N_2) \implies false$ i.e. we cannot have two nodes where one decides to commit and the other aborts.   

Instantiating a violation in an arbitrary size instance $L$, we get two nodes $a_L$ and $b_L$ such that $decide\_commit(a_L) \land decide\_abort(b_L)$ is true. Consequently, according to our strategy, we have a two node cutoff system with nodes $a_C$ and $b_C$. Because the protocol has actions that require universal quantifiers (such as $go\_commit$ which requires $\forall n. \: vote\_yes(n)$)
the sets $S$ and $A$ obtained in the static analysis require us to simulate actions of all nodes in the larger system. Therefore, according to our strategy, we map actions and state components of the rest of the nodes to one of $a_C/b_C$. This simulation strategy however does not work, because there are inherent constraints among the actions of the protocol. For example, $vote\_yes(n)$ requires $alive(n)$ and $\neg vote\_no(n)$. $vote\_no(n)$ requires $\neg vote\_yes(n)$ and $alive(n)$. Therefore, a node in the larger system cannot vote yes or vote no after failing. Similarly, a node cannot vote yes after voting no and vice versa. However, when we map multiple nodes of the larger system to the cutoff system, we can no longer maintain those constraints. For example, lets say two nodes $d_1$ and $d_2$ map to the same node $n_C$(one of $a_C$/$b_C$) in the cutoff system. Because of the $sim$ mapping, if $d_1$ performs the $fail$ action, we require $n_C$ to do the same thereby setting $alive(n_C) = false$ in the cutoff system. Now if $d_2$ performs $vote\_yes$, we can no longer perform same at the node $n_C$ because it does not satisfy the guard for the action anymore. Essentially, \textit{the side effects of some actions that we are required to simulate in the cutoff system to maintain the simulation relation are affecting the correctness of some other simulation relation clauses}

To make the simulation strategy work, we needed to the following changes
\begin{itemize}
\item We introduce a new node $f$ in the cutoff system which specifically simulates the failures of nodes in the larger system i.e. when any node fails in the larger system, we trigger the $fail(f)$ action in the cutoff system. 
\item The $vote\_yes$ action for any node in the larger system also triggers $vote\_yes$ in the node $f$. This is required to simulate a corner case behaviour where nodes can fail after voting yes. 
\item We also added an extra constraint which says that $vote\_yes(n)$ and $vote\_no(n)$ actions cannot occur together in any execution.
\end{itemize}

After adding the appropriate simulation relation clauses necessary to trigger the above actions, we were able to verify the correctness of the new cutoff instance. 

\noindent \textbf{Distributed Lock Server}: The Distributed Lock Server protocol \cite{Ivy} implements a decentralized lock service. The protocol works by maintaining an ever increasing $epoch$. A sender node can send the lock to a receiver at an epoch $e$ which is greater than the senders current epoch provided it hasn't already sent a lock at its current epoch. A receiver can receive a transfer provided that the epoch in the message $e$ is greater than its current epoch. The receive action transfers ownership to the receiver and also updates the receivers epoch to $e$. The safety property states that two different nodes cannot own the lock in the same epoch. Intuitively, the protocol works because the node with the highest epoch value owns the lock at any state and since the epoch value when transferring the lock is strictly greater than that of the sending node, the receiving node can only receive the lock at a higher epoch. 

Similar to previous protocols, we have 2 violating nodes in the larger system and correspondingly 2 cutoff system nodes. The $sim$ mapping obtained maps one of the two violating nodes directly and the rest of the nodes to the remaining cutoff system node. However, the synthesized simulation relation and lockstep fall short for two reasons
\begin{itemize}
\item Because we have many nodes of the larger system mapping to one node in the cutoff system, we cannot state that $\forall N. \: epoch_L(N) = epoch_S(sim(N))$ which contradicts one of the main correctness properties of the protocol.
We fix this by maintaining that $epoch_S(n_c) = \max\limits_{sim(n) = n_c} epoch(n)$
\item We also require to change the protocol description for the cutoff instance slightly by allowing it to accept stale transfer messages (ones with lower epochs than the current epoch of the receiver) without modifying the epoch of the receiving node. However, this new version of the protocol still maintains the same safety property.  
\end{itemize}

\noindent \textbf{Basic Key Value Store}: We also applied our approach on a simplified version of the Sharded Key Value store protocol\cite{FW19}. This version omits the logic involving sequence numbers, retransmissions, message drops and acknowledgement. The protocol has two basic actions, $reshard$ action removes a key-value pair from the table of a node and sends a $transfer\_message$ to a destination node. The $recv\_transfer\_msg$ actions uses a pending $transfer\_msg$ and installs the key-value pair in the destination node. Surprisingly, this simplified version of the protocol requires an additional tweak and does not directly synthesize a correct cutoff instance as-per our strategy. 

The violation is same as before which contains two nodes $a_L$, $b_L$ such that $table(a_L, k, v_1)$, $table(b_L, k, v_2)$. The cutoff instance therefore has two nodes $a_C$ and $b_C$. However, the generated simulation relation and lockstep fail to prove the correctness of the cutoff instance. The primary reason is that in the version of the protocol with sequence numbers, we have unique sequence numbers that mark each transfer message. Therefore, two different transfer messages are guaranteed to have two different sequence numbers. To make the simulation work, we assert a pre-condition which disallows prior states that have two different transfer messages with the same key. 
\\
\section{Related Work and Conclusion}

In the recent past, there has been a lot of interest in automated and mechanised verification of distributed protocols  \cite{FW19,MG19,MP20,PL17,Yao21,Damian19}. Ironfleet \cite{Ironfleet} and Verdi \cite{Wilcox2015VerdiAF} are some of the earliest works which are more focused towards verifying real-world implementations of distributed protocols, and typically assume that the model of the protocol works correctly. Damian et al. \cite{Damian19} use the round structure in protocols to simplify the problem and reduce the asynchrony and non-determinism, but then use standard deductive verification techniques for verifying the simplified protocol. Many of the recent approaches towards protocol verification rely on constructing and proving some form of inductive invariant. Padon et. al. \cite{Ivy} introduced the Ivy framework along with the RML language which allows a protocol developer to interactively generate an inductive invariant for verifying safety. Other approaches \cite{FW19,MG19,Yao21} have continued along this line of work, by attempting to automate the process of deriving the inductive invariant. Feldman el.al. \cite{FW19} uses the IC3/PDR approach to derive so-called phase invariants which hold for specific phases of the protocol. Both Ma et al. \cite{MG19} and Yao el al. \cite{Yao21} use a data-driven approach, whereby they execute smaller instances of the protocol to derive properties that are then generalized for arbitrary instances. While these approaches have been successful to some extent, we note that the problem of deriving inductive invariants is a fundamentally hard problem, and our work allows us to sidestep it. In fact, it could be useful to apply phase-based or data-driven approaches to the comparatively simpler problem of finding and proving a cutoff instance. In particular, the simulation and lockstep relation that we use for verifying a cutoff instance also need to be inductive in nature, but are significantly simpler in their complexity, and could be more amenable to invariant synthesis techniques.

While previous works have also attempted to use cut-off based approaches for verification  \cite{RingCutoffs,GSP,cutoffconsensus,bloem_decidability_2015}, they have mostly been limited to either a restricted class of protocols or a restricted class of specifications. Among the more recent works on parameterized cut-off based verification, Maric et al. \cite{cutoffconsensus} propose a technique to automatically determine cutoff bounds for consensus algorithms, while Jaber et. al \cite{GSP} considers protocols which use consensus as a building block, and propose a cut-off based approach to verify them. We note that none of these works actually mechanize and automate the proof that a protocol instance is actually a cutoff instance. To our best knowledge, ours is the first work that enables mechanized cut-off based verification, which is especially desirable since pen-and-paper proofs can result in errors. 

To conclude, in this work, we investigated the applicability of cutoff based verification for a variety of distributed protocols. We observe that cutoff based verification allows us to naturally sidestep the harder problem of finding inductive invariants. We identify sufficient conditions which can be used to verify that a protocol instance is indeed a cutoff instance and which can be encoded using SMT. We develop a simple static analysis-based approach to automatically synthesize the cut-off instance for many protocols. The cut-off based verification approach demonstrates how a combination of static analysis, SMT-based verification, and model checking can simplify the hard problem of protocol verification. Our experimental results indicate that cut-off results are ubiquitous and applicable for different types of protocols. Our vision is that this work can pave the way for more investigations into automating cut-off results for more complex protocols.
\nocite*{}
\bibliographystyle{splncs04}
\bibliography{biblio}

\appendix
\section{Proofs}

\textbf{Theorem 1.} For a distributed protocol $\texttt{P}$, if $\mathcal{C}$ is a cut-off instance, and $\mathcal{A}_\mathcal{C}^\texttt{P}$ is safe, then the distributed protocol $\texttt{P}$ is safe.

\begin{proof}
The proof follows directly from the definitions of a cut-off instance and safety of a distributed protocol.
\end{proof}

\textbf{Theorem 2.} Given a distributed protocol \texttt{P} and a valid interpretation $\mathcal{C}$, if for any arbitrary valid interpretation $\mathcal{L}$, there exists a simulation relation $\gamma_\mathcal{L}$ such that $(\varphi_{init} \wedge \varphi_{step} \wedge \varphi_{safety})(\gamma_\mathcal{L})$, then $\mathcal{C}$ is a cut-off instance of \texttt{P}.

\begin{proof}
To show that $\mathcal{C}$ is a cutoff instance, we need to show that if $\mathcal{C}$ is safe, then any protocol instance will be safe. We will show the contra-positive. Consider some protocol instance $\mathcal{L}$ which is not safe. Then, there exists a trace $\tau = \sigma_0^\mathcal{L} a_1^\mathcal{L} \sigma_1^\mathcal{L} \ldots a_n^\mathcal{L} \sigma_n^\mathcal{L}$ such that $\sigma_n^\mathcal{L} \not\models \Phi$, where $\Phi$ is the safety property of the protocol. We will construct by induction a trace of the cutoff instance $\tau' = \sigma_0^\mathcal{C} a_1^\mathcal{C} \sigma_1^\mathcal{C} \ldots a_m^\mathcal{C} \sigma_m^\mathcal{C}$ such that $\sigma_m^\mathcal{C} \not\models \Phi$.

\textit{Base case:} By $\varphi_{init}(\gamma_\mathcal{L})$, we know that there exists $\sigma_0^\mathcal{C}$ such that $(\sigma_0^\mathcal{L},\sigma_0^\mathcal{C}) \in \gamma_\mathcal{L}$.

\textit{Inductive case:} Consider some state $\sigma_i^\mathcal{L}$ in the trace $\tau$ such that there exists $\sigma_j^\mathcal{C}$ in the trace $\tau'$ and $(\sigma_i^\mathcal{L},\sigma_j^\mathcal{C}) \in \gamma_\mathcal{L}$. Since $(\sigma_i^\mathcal{L}, a_{i+1}^\mathcal{L}, \sigma_{i+1}^\mathcal{L}) \in \delta_\mathcal{L}$, by $\varphi_{step}(\gamma_\mathcal{L})$, there would exists a finite sequence of transitions in $\delta_\mathcal{C}$ beginning from $\sigma_j^\mathcal{C}$ and ending in some $\sigma_k^\mathcal{C}$. We append these transitions to the end of the trace $\tau'$ constructed so far.

Hence, at the end, we must have $(\sigma_n^\mathcal{L},\sigma_m^\mathcal{C}) \in \gamma_\mathcal{L}$. By $\varphi_{safety}(\gamma_\mathcal{L})$, we must have $\sigma_m^\mathcal{C} \not\models \Phi$, thus proving the result.
\end{proof}

\textbf{Lemma 1.} Given a distributed protocol \texttt{P} and a valid interpretation $\mathcal{C}$, if for any arbitrary valid interpretation $\mathcal{L}$, there exists a simulation relation $\gamma_\mathcal{L}$ such that $(\varphi_{init} \wedge \varphi^{first}_{step} \wedge \varphi_{safety})(\gamma_\mathcal{L})$, then $\mathcal{C}$ is a cut-off instance of \texttt{P}.

\begin{proof}
The proof is the same as the proof for Theorem 2, except that we consider a minimal violating trace of the protocol instance $\mathcal{L}$, such that only the final state in the trace does not satisfy the safety property.
\end{proof}
\end{document}